\documentclass[english,aps,prstper,reprint,showpacs,titlepage,longbibliography]{revtex4-2}   

\usepackage[T1]{fontenc}	
\usepackage{geometry}		
\usepackage{graphicx}
\usepackage[above,below]{placeins}	
\usepackage{times}
\usepackage[most]{tcolorbox}

\usepackage{hyperref}  
\hypersetup{colorlinks=true,urlcolor=blue,citecolor=blue,linkcolor=blue}   
\usepackage{enumitem}          
\setlist{nosep}                 

\begin{document}

\begin{titlepage}

\title{How Well Do AI Systems Solve AP Physics? A Comparative Evaluation of Large Language Models on Algebra-Based Free Response Questions}

 \author{Bilas Paul}
\email[Please address correspondence to ]{palb@farmingdale.edu}
\affiliation{SUNY Farmingdale State College, Farmingdale, NY 11735, USA}
\author{Jashandeep Kaur}
\affiliation{SUNY Farmingdale State College, Farmingdale, NY 11735, USA}
\author{Shantanu Chakraborty}
 \affiliation{Valdosta State University, Valdosta, GA 31698, USA} 
\author{Shruti Shrestha}
 \affiliation{Penn State Brandywine, Media, PA 19063, USA}

\begin{abstract}
\label{abstract}
\noindent The rapid advancement of large language models (LLMs) has generated growing interest in their potential role in physics education and assessment, yet a focused evaluation of their performance on multi-faceted, free-response physics problems remains underexplored. In this study, we systematically evaluate the performance of four widely accessible AI systems—ChatGPT 4.1 mini, Gemini 2.5 Flash, Claude 4.0 Sonnet, and DeepSeek R1—on algebra-based Advanced Placement (AP) Physics 1 and AP Physics 2 free-response questions administered between 2015 and 2025. Model-generated solutions were produced under standardized exam-style prompting and evaluated by three independent physics experts using official College Board scoring guidelines.  All models achieved relatively high mean scores (82–92\%), indicating strong capability in structured algebraic problem solving. However, substantial year-to-year variability was observed, particularly for AP Physics 1, where statistical testing revealed no consistent performance hierarchy among models. In contrast, AP Physics 2 results showed statistically significant differences, with Gemini and DeepSeek demonstrating more consistent performance than Claude, and numerically higher performance than ChatGPT, though the latter comparisons did not reach statistical significance.  A qualitative analysis revealed recurring error patterns across all models, including misinterpretation of diagrams and graphs, incorrect graph construction, incorrect reasoning about vector direction, circuit topology errors, partial and misleading qualitative explanations, and difficulties applying three-dimensional concepts such as the right-hand rule. These findings suggest that while contemporary AI systems can effectively support routine physics problem solving, they remain limited in tasks requiring spatial reasoning, visual interpretation, and conceptual integration. The results highlight both the instructional potential and current pedagogical limitations of AI-assisted learning tools in physics education.

\end{abstract}

\maketitle
\end{titlepage}

\section{Introduction }
\label{intro}
The rapid advancement of large language models (LLMs) such as ChatGPT, Claude, and Gemini has sparked growing interest in their potential applications within education, particularly within science, technology, engineering, and mathematics (STEM) disciplines. As these AI systems become increasingly accessible, educators and researchers are keen to understand both their capabilities and limitations in supporting student learning and assessment~\cite{Wang2024ChatGPTScienceEngineering, Liang2023ChatGPTPhysics, WangFan2025EffectChatGPT}. One area of particular interest is their performance  on standardized assessments, such as the Advanced Placement (AP) Physics exams, which are widely recognized for their rigorous evaluation of students’ conceptual understanding, problem-solving skills, and ability to communicate scientific reasoning. 

The AP Physics 1 and AP Physics 2 (both algebra-based) exams, designed by the College Board, include a substantial free-response section that challenges students to apply physics principles in a variety of contexts. These free-response questions (FRQs) require not only quantitative calculations but also qualitative explanations, experimental design, and the translation between multiple representations (e.g., graphs, diagrams, and equations). The complexity and open-ended nature of these questions make them a robust benchmark for evaluating the reasoning and problem-solving abilities of AI systems.

Recent research has systematically examined the performance of LLMs such as ChatGPT, Claude, and Gemini on problem-solving and reasoning tasks. Studies indicate that while these models can achieve high accuracy on well-specified, textbook-style problems, their performance drops significantly on open-ended or under-specified questions that require deeper conceptual understanding, creative reasoning, or the ability to make reasonable assumptions about missing information~\cite{Wang2024ChatGPTScienceEngineering, Liang2023ChatGPTPhysics}.  Common failure modes identified in the literature include difficulties in constructing accurate models of physical scenarios, numerical inaccuracies, and misalignment between problem statements and relevant physics principles. Furthermore, while AI-generated responses often produce grammatically correct and assertive explanations, they may lack metacognitive awareness and fail to notice inconsistencies in its own reasoning~\cite{SIRNOORKAR2024100318}.

Comparative analyses of leading LLMs have provided valuable insights into the differentiated reasoning patterns and performance profiles across STEM-related tasks.  Empirical analyses indicate that while ChatGPT tends to perform consistently on structured, calculation-based questions, Claude exhibits stronger multi-step reasoning and reflective coherence, particularly in open-ended scenarios that require logical explanation and contextual inference~\cite{Srinivasa2025TutorBench}. In contrast, Gemini’s multimodal integration of text and visual inputs provides advantages in interpreting graphs, diagrams, and symbolic representations, which are central to physics education and AP-style free-response assessments.  Furthermore, research on automated grading systems has demonstrated that advanced prompt engineering strategies, including chain-of-thought reasoning, can significantly enhance model performance in assessing complex calculation problems~\cite{educsci15020116}. Collectively, these studies underscore the importance of cross-model comparisons for understanding how architectural and training differences influence the reasoning, consistency, and pedagogical suitability of AI systems in educational contexts.

While there is a growing body of research examining the use of AI in education,  a focused, granular analysis of their performance on the specific, multi-faceted FRQs of the AP Physics exams remains underexplored.
Addressing this gap is essential for informing both the responsible integration of AI tools in physics education and the ongoing development of more robust and pedagogically aligned AI systems~\cite{WangFan2025EffectChatGPT, Cong-Lem03042025}.  In this paper, we systematically evaluate the performance of several large language model–based AI platforms on a representative set of algebra-based AP Physics FRQs. Our primary objective is to identify the types of questions and reasoning tasks where these systems demonstrate strength, as well as those that reveal persistent limitations. By analyzing AI-generated responses against official scoring rubrics and established assessment frameworks, we aim to provide actionable insights for educators, researchers, and developers. These insights seek to support the responsible integration of AI in physics education while remaining cognizant of its current cognitive and pedagogical constraints.
 

 \section{Methods and Materials}
The study utilized FRQs from the AP Physics 1 and AP Physics 2 exams administered by the College Board between 2015 and 2025, excluding 2020 due to the absence of a standardized examination format during the COVID-19 pandemic. The selected questions are publicly available with official scoring guidelines and are rigorously designed to assess conceptual understanding, problem-solving skills, and scientific reasoning. They are widely recognized as reliable benchmarks for evaluating cognitive and procedural competencies in physics education. Together, these questions span a diverse range of physics topics, including but not limited to kinematics, dynamics, energy, momentum, rotational motion, electricity and circuits, waves, and modern physics.  Both exams are algebra-based and include problems that require students to engage in multiple forms of scientific reasoning: interpreting and creating diagrams, performing quantitative calculations with appropriate units, designing and analyzing experiments, constructing and interpreting graphs, and providing qualitative explanations that demonstrate conceptual understanding. This multi-faceted assessment structure makes AP Physics FRQs particularly well-suited for evaluating the breadth and depth of LLM capabilities in authentic physics problem-solving contexts.

Four distinct LLMs were selected for generating the free-response solutions: ChatGPT 4.1 mini (Open AI), Gemini 2.5 Flash (Google DeepMind), Claude 4.0 Sonnet (Anthropic), and Deepseek R1 (Deepseek AI). These models were chosen because they represent leading commercial LLM platforms with demonstrated capabilities in STEM reasoning tasks and are publicly accessible to educators and students. The specific model versions were carefully documented to ensure replicability and to account for the rapid evolution of these systems through ongoing updates and fine-tuning.  Each model was accessed through its respective official web interface during the period of October--December 2025. 
All interactions were conducted using the free-tier or standard-access versions of each platform to reflect the typical user experience available to educators and students. No specialized prompting techniques (e.g., chain-of-thought scaffolding, few-shot examples, or iterative refinement) were employed beyond the standardized instructional prompt described below, ensuring that the evaluation reflected the models' base-level performance rather than optimized configurations.

To maintain uniform testing conditions and minimize prompt bias across models, a standardized instructional prompt was developed and applied consistently to all questions and all AI systems. The prompt was designed to emulate the instructions and expectations of the actual AP Physics exam environment, framing the AI as a student test-taker rather than an instructional assistant. This approach ensured that model outputs were directly comparable to human student responses and could be fairly evaluated using the official College Board scoring rubrics. The complete standardized prompt is reproduced below:

\begin{tcolorbox}[colback=violet!10, colframe=violet!40, title=Standardized AI Prompt]
You are a high school student taking the AP Physics exam. You are given a free-response question that includes both a textual description and a diagram. Your task is to solve the problem as a student would during the exam, providing a clear, concise, and accurate response.\\[4pt]
\textbf{Instructions:}
\begin{enumerate}
\item[1.] Use only the information given in the question (including the diagram).
\item[2.] Answer step-by-step, showing all reasoning, formulas, and calculations clearly, with appropriate units.
\item[3.] Apply relevant physics concepts based on the problem.
\end{enumerate}
\end{tcolorbox}
\begin{tcolorbox}[colback=violet!10, colframe=violet!40]
\begin{enumerate}
\item[4.] Reference the diagram explicitly when it informs your solution (e.g., angles, directions, or quantities).
\item[5.] Keep your response focused and to the point, just like in an actual exam setting. Avoid unnecessary details.
\item[6.] If a question asks for an explanation, write it in complete sentences.
\item[7.] If a question asks for a calculation, show the formula, plug in values, and give a final boxed answer with units.
\item[8.] If the question asks you to sketch or plot a graph, or draw a diagram, describe clearly what it would look like:
\begin{enumerate}
\item For graphs: label axes, indicate shape/trend, and note key points (e.g., intercepts, maxima/minima).
 \item For diagrams: describe objects, forces, directions, labels, and key features.
\end{enumerate}
 \item[9.] Do not include anything beyond what a student would write in a test booklet.
 \item[10.] Do not say you are an AI or reference being an AI.
\end{enumerate}
\end{tcolorbox}

Each question was submitted sequentially to all platforms, and model outputs were recorded in full without post-processing or modification. To ensure consistent and authentic application of the prompt, all responses were generated by a research assistant who had completed two semesters of algebra-based  physics (College Physics I and College Physics II). The assistant was instructed to carefully follow the standardized prompt when interacting with each platform, ensuring consistency in question input, adherence to instructions, and verification that the model responses fully addressed the given tasks. This procedure minimized human bias while maintaining a level of contextual understanding consistent with realistic exam conditions.

The AI-generated responses were assessed using the official College Board scoring guidelines (rubrics) corresponding to each year’s AP Physics FRQs. These rubrics provide detailed performance criteria for conceptual reasoning, numerical accuracy, correct application of physics principles, and clarity of explanation.   Each FRQ typically awards 7--12 points distributed across multiple subparts, with partial credit available for partially correct reasoning or calculations. 
To establish inter-rater reliability, all responses were independently scored by three expert raters from three different US institutions, each having advanced degrees in physics and extensive experience in teaching introductory college-level physics. Inter-rater agreement was quantified using the Intraclass Correlation Coefficient [ICC(2,k)]~\cite{Koo2016Guideline} and Cronbach's alpha~\cite{Tavakol2011Cronbach}. The final score for each response was calculated as the mean across the three raters, which accounts for variations in scoring interpretation while reflecting overall expert judgment.

Statistical analyses were conducted using Python 3.10 with SciPy, NumPy, and pandas libraries.  Performance consistency across exam years was quantified using the coefficient of variation (CV = $\sigma/\mu \times 100\%$). To compare model performance, we employed the Friedman test (non-parametric repeated-measures analysis) with exam year as the repeated measure~\cite{Friedman1937Use}. For significant omnibus results, post-hoc pairwise comparisons were conducted using Wilcoxon signed-rank tests with Bonferroni correction ($\alpha_{\text{corrected}} = 0.05/6$)~\cite{Wilcoxon1945Individual}. Effect sizes were calculated using Cohen's $d$~\cite{Cohen1988Statistical} and Kendall's $W$ (coefficient of concordance)~\cite{Kendall1948Rank}. Levene's test assessed homogeneity of variance across models~\cite{Levene1960Robust}. All tests were two-tailed with $\alpha = 0.05$. Complete analysis code and data are available in supplementary materials.


\section{Results}
\label{results}

To establish the validity of our multi-rater scoring approach, we first assessed the agreement among the three independent raters using the Intraclass Correlation Coefficient [ICC(2,k)] and Cronbach's alpha. These measures quantify the consistency and reliability of the scoring process, which is essential for interpreting subsequent comparisons of AI model performance.

For AP Physics 1, inter-rater reliability was excellent across all four AI models, substantially exceeding the threshold for excellent agreement (ICC $\geq$ 0.75). ICC values ranged from 0.909 to 0.935, with DeepSeek demonstrating the highest agreement (ICC = 0.935, 95\% CI [0.849, 0.985]). Cronbach's alpha values ranged from 0.974 to 0.981, indicating that the three raters applied the scoring rubrics with exceptional consistency across all 10 exam years. On the other hand, for AP Physics 2, inter-rater reliability showed greater variability across models but remained generally strong. ChatGPT and Claude exhibited excellent reliability (ICC = 0.896 and 0.828, respectively), while Gemini demonstrated good reliability (ICC = 0.715) and DeepSeek showed fair reliability (ICC = 0.589). Despite the wider ICC range, Cronbach's alpha remained acceptable to excellent (0.829--0.971), indicating that while absolute score assignments varied more for some models, raters maintained consistent patterns in their scoring judgments. The lower ICC values for certain models in Physics 2 likely reflect the increased complexity of evaluating open-ended problems requiring multi-step reasoning, experimental design, and integration of multiple physics concepts. Table~\ref{tab:icc_reliability} presents complete reliability statistics.

\begin{table}[h]
\centering
\caption{Inter-Rater Reliability Statistics for AI Model Scoring}
\label{tab:icc_reliability}
\begin{tabular}{llccc}
\hline
\textbf{Subject} & \textbf{Model} & \textbf{ICC(2,k)} & \textbf{95\% CI} & \textbf{Cronbach's $\alpha$} \\
\hline
Physics 1 & ChatGPT  & 0.912 & [0.802, 0.979] & 0.974 \\
          & Gemini   & 0.917 & [0.808, 0.980] & 0.975 \\
          & Claude   & 0.909 & [0.805, 0.980] & 0.975 \\
          & DeepSeek & 0.935 & [0.849, 0.985] & 0.981 \\
\hline
Physics 2 & ChatGPT  & 0.896 & [0.771, 0.975] & 0.969 \\
          & Gemini   & 0.715 & [0.491, 0.932] & 0.912 \\
          & Claude   & 0.828 & [0.785, 0.977] & 0.971 \\
          & DeepSeek & 0.589 & [0.250, 0.873] & 0.829 \\
\hline
\end{tabular}
\begin{flushleft}
\small
\textit{Note:} ICC(2,k) = Intraclass Correlation Coefficient (two-way random effects, average measures). Interpretation: $<0.40$ = poor, $0.40-0.59$ = fair, $0.60-0.74$ = good, $\geq 0.75$ = excellent. Cronbach's $\alpha > 0.70$ is considered acceptable; $> 0.80$ is good; $> 0.90$ is excellent.
\end{flushleft}
\end{table}

The excellent reliability for Physics 1 (all ICC $> 0.90$) and generally strong reliability for Physics 2 (ICC = 0.589--0.896) provide a solid foundation for comparing AI model performance. All subsequent analyses use mean scores across the three raters for each model-year combination.




\begin{table*}[ht!]
\centering
\caption{Descriptive Statistics for AI Model Performance}
\label{tab:descriptive_stats}
\begin{tabular}{llcccccc}
\hline
\textbf{Subject} & \textbf{Model} & \textbf{Mean (\%)} & \textbf{SD} & \textbf{Median (\%)} & \textbf{IQR} & \textbf{Range} & \textbf{CV (\%)} \\
\hline
Physics 1 & ChatGPT  & 86.6 & 10.5 & 92.1 & [75.3, 95.2] & [73.0, 97.8] & 12.1 \\
          & Gemini   & 88.9 & 10.9 & 90.9 & [80.4, 98.9] & [72.6, 100.0] & 12.3 \\
          & Claude   & 91.7 & 8.9  & 94.9 & [92.8, 96.3] & [71.9, 100.0] & 9.7 \\
          & DeepSeek & 89.7 & 10.6 & 93.7 & [80.3, 97.4] & [72.0, 100.0] & 11.8 \\
\hline
Physics 2 & ChatGPT  & 82.5 & 10.4 & 81.1 & [73.8, 91.3] & [70.0, 97.7] & 12.6 \\
          & Gemini   & 91.2 & 5.7  & 91.2 & [86.0, 95.9] & [83.2, 99.3] & 6.3 \\
          & Claude   & 84.1 & 6.3  & 84.9 & [81.0, 88.4] & [73.1, 92.9] & 7.5 \\
          & DeepSeek & 92.0 & 4.3  & 92.0 & [89.5, 95.5] & [84.1, 97.8] & 4.7 \\
\hline
\end{tabular}
\begin{flushleft}
\small
\textit{Note:} SD = standard deviation; IQR = interquartile range (Q1 to Q3); CV = coefficient of variation. All statistics calculated using mean scores across three independent raters.
\end{flushleft}
\end{table*}

Table~\ref{tab:descriptive_stats} presents comprehensive descriptive statistics for each AI model's performance across the 10 exam years, calculated using mean scores across the three independent raters. All four models demonstrated strong overall performance on both subjects, with mean scores consistently exceeding 82\%. However, the models exhibited markedly different performance profiles when examining central tendency, dispersion, and consistency measures.

For AP Physics 1, the four models occupied a narrow performance band, with means ranging from 86.6\% (ChatGPT) to 91.7\% (Claude)—a span of only 5.1 percentage points. This tight clustering suggests that current frontier AI systems have achieved comparable overall capability on algebra-based mechanics problems. However, substantial standard deviations (8.9--10.9\%) and coefficients of variation (9.7--12.3\%) reveal considerable year-to-year fluctuation, with all models exhibiting similar levels of variability across the decade of exams. AP Physics 2 revealed a more differentiated performance landscape. Gemini (91.2 $\pm$ 5.7\%) and DeepSeek (92.0 $\pm$ 4.3\%) achieved the highest mean scores, outperforming Claude (84.1 $\pm$ 6.3\%) by approximately 7 percentage points  and ChatGPT (82.5 $\pm$ 10.4\%) by approximately 9 percentage points.  Critically, Physics 2 was characterized by substantially lower variability for three models (CV = 4.7--7.5\% for Gemini, Claude, and DeepSeek), with DeepSeek and Gemini exhibiting exceptional consistency (CV = 4.7\% and 6.3\%). ChatGPT showed notably greater year-to-year fluctuation (CV = 12.6\%), more than double that of the most consistent models.

\begin{figure*}[htb!]
\centering
\includegraphics[width=0.95\textwidth]{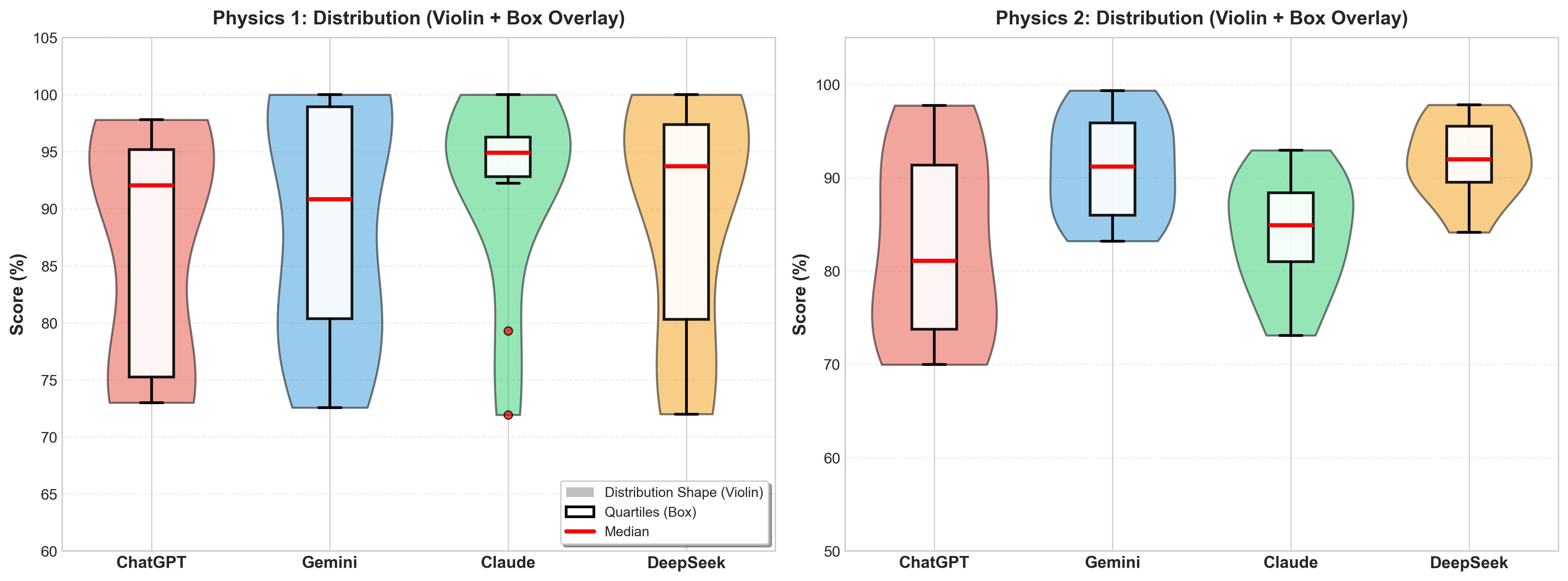}
\caption{Performance distributions combining violin plots (probability density) and box plots (statistical summaries). Violin shapes reveal distribution characteristics: wider regions indicate score concentrations while narrow regions show sparse data. Overlaid box plots display median (thick red line), interquartile range (box), whiskers (range), and outliers (red dots). (left) Physics 1 shows broader distributions with some bimodality, particularly for ChatGPT and Gemini, indicating inconsistent performance across exam years. (right) Physics 2 displays contrasting patterns: Gemini and DeepSeek exhibit narrow, sharply peaked distributions indicating consistent high performance, while ChatGPT shows greater spread with an outlier at 100\% (2025 exam), illustrating capacity for perfect performance but lower typical scores.}
\label{fig:distributions}
\end{figure*}

Figure~\ref{fig:distributions} complements these statistics by visualizing complete performance distributions through overlaid violin and box plots. The violin shapes (colored, semi-transparent regions) display probability density—width indicates concentration of scores—revealing whether distributions are uniform or peaked. Overlaid box plots (white boxes with black edges) provide precise statistical summaries: median (thick red line), interquartile range (box height), range (whiskers), and outliers (red dots).

For Physics 1 (Figure~\ref{fig:distributions}, left), violin plots display relatively broad, sometimes bimodal distributions, particularly for ChatGPT and Gemini. ChatGPT's violin shows substantial density around both 75\% and 95\%, suggesting dichotomous performance rather than consistent middle-range scores. Box plots confirm this spread: ChatGPT's IQR spans 75.3--95.2\%, while Claude shows the narrowest box (IQR: 92.8--96.3\%) despite similar overall ranges across models. The similar distribution ranges across all four models reflect the comparable coefficients of variation (9.7--12.3\%).  Physics 2 distributions (Figure~\ref{fig:distributions}, right) tell a different story. Gemini and DeepSeek exhibit narrow, sharply peaked violins, indicating tightly clustered scores—a hallmark of consistent, reliable performance. Their box plots show compact IQRs spanning only 6--10 percentage points. ChatGPT's violin remains broader and flatter, with a wider IQR (73.8--91.3\%) and lower median (81.1\%). Notably, ChatGPT displays an outlier at 97.7\% (2025 exam), illustrating that while capable of perfect performance under specific conditions, typical performance falls substantially lower. This visual contrast between ChatGPT's broad distribution and the peaked distributions of Gemini and DeepSeek mirrors the marked difference in their coefficients of variation (12.6\% vs. 4.7--6.3\%).

\begin{figure*}[htb!]
\centering
\includegraphics[width=0.95\textwidth]{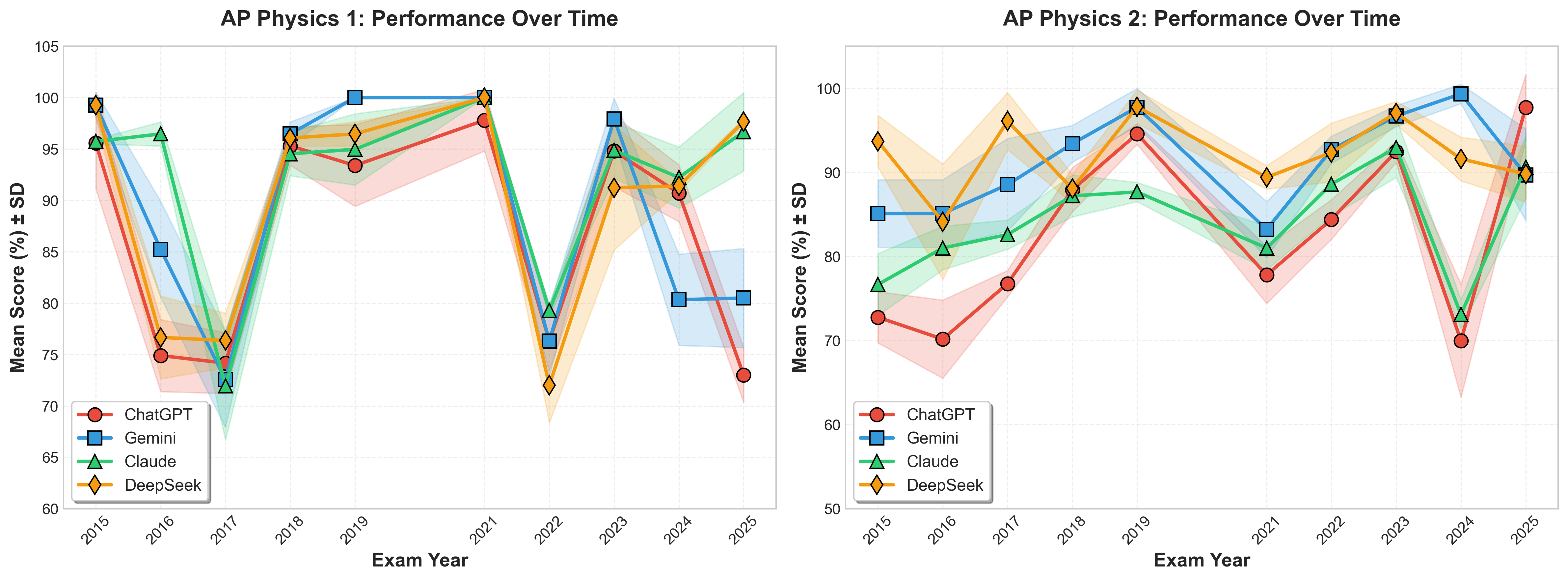}
\caption{Temporal trends in AI model performance with inter-rater variability. Mean scores (solid lines with markers) and standard deviation bands (shaded regions) across three independent raters for (left) AP Physics 1 and (right) AP Physics 2. The shaded regions illustrate scoring consistency, with narrower bands indicating higher inter-rater agreement. Substantial year-to-year fluctuations are evident, with certain exam years (e.g., 2021) eliciting consistently high performance across all models while others (e.g., 2017, 2022) proved universally challenging. The crossing and diverging trajectories indicate that model rankings are not stable but rather depend on specific exam characteristics.}
\label{fig:performance_time}
\end{figure*}

Figure~\ref{fig:performance_time} displays the temporal evolution of model performance across the decade of exams studied, with shaded regions representing $\pm$1 standard deviation across the three raters. This visualization reveals substantial year-to-year variability, exam-specific difficulty patterns, and model-dependent trajectories.

For Physics 1 (Figure~\ref{fig:performance_time} left), model trajectories intersect frequently across the 10-year span, illustrating the instability of performance rankings. The 2021 exam stands out as an anomaly, eliciting near-perfect or perfect performance from all models (97.8--100\%), suggesting problem characteristics—perhaps more straightforward application of formulas or reduced conceptual complexity—that were particularly amenable to current AI architectures. In contrast, the 2022 exam (mean scores: 72.0--79.3\%) and 2017 exam (71.9--76.4\%) proved more challenging across all models. Within-model variability was dramatic: Claude's performance ranged from 71.9\% (2017) to 100\% (2021), a 28.1 percentage point swing, while similar patterns characterized all other models. The 2025 exam revealed the most pronounced model divergence, with scores spanning from 73.0\% (ChatGPT) to 97.7\% (DeepSeek), underscoring that exam difficulty is not universal but rather model-dependent. Physics 2 temporal patterns (Figure~\ref{fig:performance_time}, right) exhibited similar year-to-year variability but with clearer model differentiation. Gemini and DeepSeek maintained consistently elevated performance across most years, rarely falling below 83\%, 
while ChatGPT exhibited greater volatility, including a notable dips to 70.2\% in 2016 and 70.0\% in 2024. The 2024 exam revealed particularly stark performance divergence: Gemini achieved 99.3\% while ChatGPT and Claude scored substantially lower at 70.0\% and 73.1\%, respectively, with DeepSeek at 91.6\%.  The shaded SD bands provide additional insight into scoring reliability. Narrower bands (e.g., Physics 2, Gemini and DeepSeek) indicate high inter-rater agreement, while wider bands reflect greater scoring subjectivity. This validates our multi-rater approach.

The temporal instability evident in Figure~\ref{fig:performance_time} motivated formal statistical testing to determine whether observed differences among models were systematic or attributable to year-specific variation. We employed the Friedman test—a non-parametric alternative to repeated-measures ANOVA—treating exam year as the repeated measure to test whether models differ systematically in performance rankings.

\begin{table}[htb!]
\centering
\caption{Statistical Comparisons for AP Physics 1 and Physics 2}
\label{tab:posthoc_p2}
\begin{tabular}{llcc}
\hline
\textbf{Subject} & \textbf{Comparison} & \textbf{$p$ (corrected)} & \textbf{Sig.?}  \\
\hline
\multicolumn{4}{l}{\textit{Physics 1: Friedman} $\chi^2(3) = 5.46$, $p = 0.141$, $W = 0.182$} \\
\multicolumn{4}{l}{(No post-hoc tests conducted due to non-significant omnibus result)} \\
\hline
\multicolumn{4}{l}{\textit{Physics 2: Friedman} $\chi^2(3) = 15.96$, $p = 0.0012$, $W = 0.532$} \\
\hline
Physics 2 & ChatGPT vs. Gemini   & 0.117 & No   \\
          & ChatGPT vs. Claude   & 1.000 & No   \\
          & ChatGPT vs. DeepSeek & 0.082 & No   \\
          & Gemini vs. Claude    & 0.023 & Yes  \\
          & Gemini vs. DeepSeek  & 1.000 & No   \\
          & Claude vs. DeepSeek  & 0.023 & Yes  \\
\hline
\end{tabular}
\begin{flushleft}
\small
\textit{Note:} Bonferroni-corrected $\alpha = 0.0083$ for 6 comparisons. Only Gemini vs. Claude achieved statistical significance after correction.
\end{flushleft}
\end{table}

For AP Physics 1, the Friedman test revealed no statistically significant differences among the four AI models ($\chi^2(3) = 5.46$, $p = 0.141$, Kendall's $W = 0.182$). The weak concordance coefficient ($W = 0.182$) indicates that model rankings varied substantially across the 10 exam years. Kendall's $W$ ranges from 0 (complete disagreement) to 1 (perfect agreement); a value of 0.182 suggests that less than 20\% of ranking variance is attributable to consistent model differences, with the remainder due to year-specific factors. In contrast, AP Physics 2 revealed statistically significant differences among models ($\chi^2(3) = 15.96$, $p = 0.0012$, Kendall's $W = 0.532$). The moderate concordance coefficient ($W = 0.532$) indicates that approximately 53\% of ranking variance is attributable to consistent model differences, with 47\% still due to year-specific variation. While substantial year-to-year variability persists, Physics 2 exhibits a more stable performance hierarchy than Physics 1.

\begin{figure*}[htb!]
\centering
\includegraphics[width=0.95\textwidth]{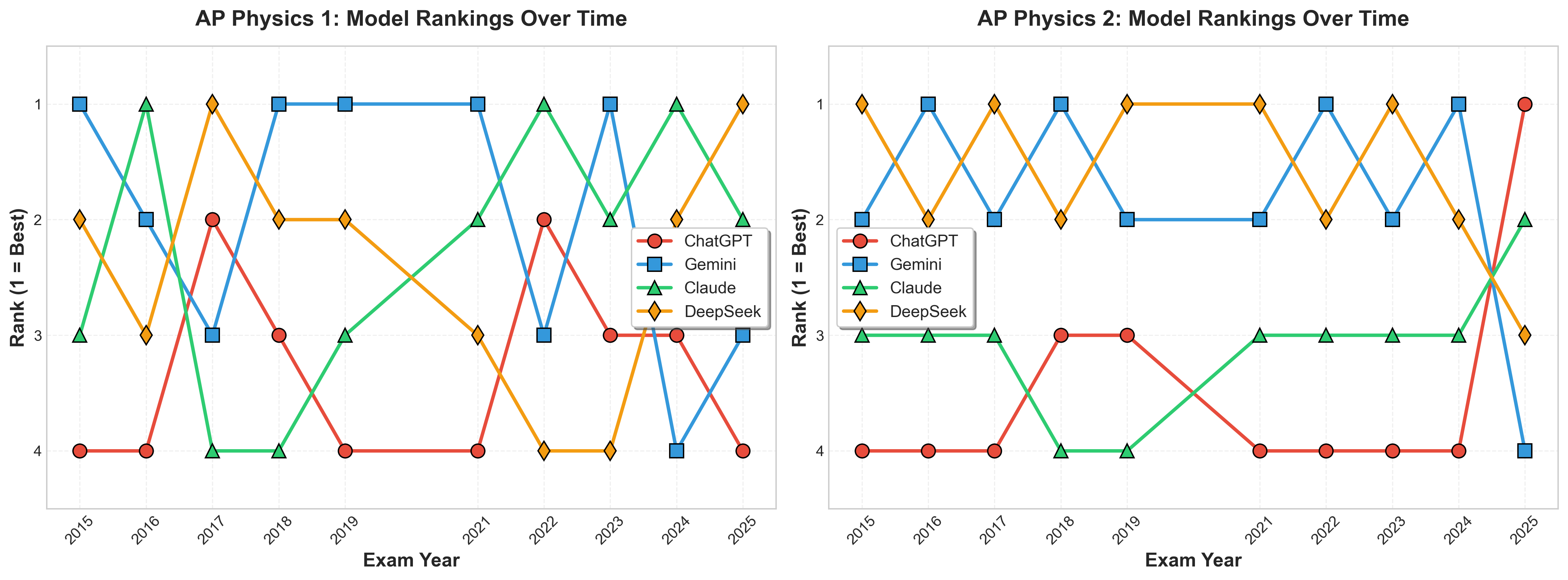}
\caption{Stability of model rankings across exam years, directly visualizing the Kendall's $W$ concordance findings. Each line traces a single model's rank (1 = highest performance, 4 = lowest) across the 10 exam years for \textbf{(Left)} AP Physics 1 and \textbf{(Right)} AP Physics 2. In Physics 1, frequent crossovers and rank reversals illustrate the low concordance ($W = 0.182$), with all models occupying each rank position multiple times and no consistent performance hierarchy emerging. Physics 2 shows more stable rankings ($W = 0.532$), with Gemini and DeepSeek predominantly occupying ranks 1-2, while ChatGPT more frequently ranks lower, consistent with the significant Friedman test and post-hoc findings.}
\label{fig:rankings}
\end{figure*}

Post-hoc pairwise comparisons using Wilcoxon signed-rank tests with Bonferroni correction ($\alpha_{\text{corrected}} = 0.05/6 \approx 0.0083$) identified two statistically significant pairwise differences for Physics 2: Gemini significantly outperformed Claude ($p = 0.023$, mean difference = 7.0 percentage points), and DeepSeek significantly outperformed Claude ($p = 0.023$, mean difference = 7.9 percentage points). Table~\ref{tab:posthoc_p2} presents complete results. The post-hoc results reveal a clear pattern: Claude performed significantly worse than both top-performing models (Gemini and DeepSeek) in Physics 2. While ChatGPT also showed numerically lower performance compared to Gemini (8.7 percentage points) and DeepSeek (9.5 percentage points), these differences did not achieve statistical significance  ($p = 0.117$ and $p = 0.082$, respectively). This pattern reflects the inherent tension in small-sample studies ($n = 10$ years) between practical performance differences and statistical power. The substantial year-to-year variability reduces statistical confidence, particularly for comparisons involving ChatGPT, which exhibited the highest variability (CV = 12.8\%). Conservative interpretation suggests that Gemini and DeepSeek hold demonstrable advantages over Claude in Physics 2, while comparisons involving ChatGPT remain ambiguous given the available data. Figure~\ref{fig:rankings} makes the ranking instability explicit by plotting each model's rank (1 = highest, 4 = lowest) across all exam years.

Figure~\ref{fig:rankings} (left) confirms the rank instability in Physics 1:  while ChatGPT never achieved top ranking, the other three models frequently alternated in first position, with Gemini ranking first most frequently (5 times), followed by Claude (3 times) and DeepSeek (2 times), though these differences were not statistically significant. This volatility, quantified by Kendall's $W = 0.182$, suggests that Physics 1 exam characteristics vary substantially in ways that differentially challenge different AI architectures. The frequent crossovers visible in this figure provide the visual counterpart to the non-significant Friedman test ($p = 0.141$). In contrast, Figure~\ref{fig:rankings} (right) reveals that Gemini and DeepSeek predominantly occupied the top two positions in Physics 2, while Claude and ChatGPT more consistently ranked lower. This relative stability, quantified by Kendall's $W = 0.532$ and visualized through fewer rank crossovers, provides the foundation for the significant Friedman test ($p = 0.0012$) and supports the post-hoc finding of both Gemini and DeepSeek significantly outperformed Claude.

\subsection{Common Errors in AI-Generated Responses}
Beyond overall performance scores, a qualitative analysis of AI-generated responses revealed recurring and systematic categories of error across all four models. These errors reflect fundamental limitations in how current LLMs process and reason about physics problems—particularly those that require spatial reasoning, visual interpretation, and consistent application of physical principles. These patterns complement the quantitative findings reported above: the same exam years that produced low scores (e.g., 2017, 2022 for Physics 1; 2016, 2024 for Physics 2) were disproportionately represented among the error instances catalogued below, suggesting that score variability is not random but reflects specific, identifiable reasoning failures. The most common error patterns are summarized below.

\begin{enumerate}
\item \textbf{Diagram Interpretation Errors:} Diagram misinterpretation represented one of the most frequent failure modes. LLMs often misidentified positions, motion states, or geometric relationships shown in figures, leading to fundamentally incorrect physical analysis.  

For example, in a scenario  (AP Physics 1, 2017) with two blocks sliding down frictionless ramps of different shapes but identical heights, several AI-generated responses incorrectly concluded that both blocks reached the floor simultaneously because they left the table from the same height. The models failed to recognize that differences in ramp steepness affect acceleration along the track and therefore the launch time from the table.  Similarly, when analyzing a transverse wave on a string (AP Physics 1, 2018) with two painted dots (P at a trough antinode, Q at a node), all four LLMs failed to correctly identify the direction of instantaneous acceleration for dot P. They did not recognize that a point at the trough of a wave has an acceleration vector pointing toward the equilibrium position.  In another scenario (AP Physics 2, 2016), two charged spheres with electrical potential isolines (10V differences) were shown. Despite explicit information about the potential difference between isolines, all four models failed to predict the correct polarity of two charged spheres surrounded by equipotential lines—a task requiring precise spatial interpretation of the isoline distribution. Across multiple FRQs, errors originated from incomplete extraction of spatial information embedded in diagrams rather than incorrect use of physics formulas.

\item \textbf{Graph Interpretation and Construction Errors:} LLMs frequently misread graphical data or failed to use numerical values explicitly provided in graphs. Instead of extracting quantitative information, models often substituted arbitrary values or relied solely on symbolic expressions. A representative example occurs in AP Physics 2 (2021), where a pressure–volume (PV) diagram was used to determine internal energy change and work done during thermodynamic processes. All evaluated models failed to read correct pressure and volume coordinates from the graph, producing inconsistent numerical results and incorrect conclusions regarding energy transfer between gas samples. This foundational error also propagated to the qualitative portion of the problem: none of the models correctly identified the temperatures corresponding to involved states, and consequently all four LLMs incorrectly predicted the direction of heat transfer when the two gas samples were brought into thermal contact. Similar issues appeared in kinematics problems requiring values from velocity–time graphs (AP Physics 1, 2024), where incorrect data extraction propagated through subsequent calculations. These findings indicate limitations in translating graphical representations into quantitative reasoning steps.

When FRQs required students to construct graphs, LLM responses frequently lacked essential physical features such as equilibrium position, amplitude, scaling, or reference points. In AP Physics 1 (2022), involving oscillations of a mass–spring system after an increase in spring constant, AI-generated responses correctly noted qualitative changes in frequency but failed to describe how the graph should be drawn. Key requirements—including unchanged amplitude and shifted equilibrium representation—were omitted. In several energy-bar diagrams, models also neglected height-dependent energy changes, demonstrating weak spatial reasoning linking graphical representations to physical geometry. Overall, LLMs tended to describe trends verbally rather than demonstrate procedurally correct graph construction.

\item \textbf{Direction-Related Errors:} Incorrect identification of vector directions was another persistent issue. Many AP Physics FRQs rely critically on directional reasoning involving forces, electric fields, or motion.  For instance, in AP Physics 1 (2022), involving gravitational interactions between two moons and a planet aligned along an axis, AI responses incorrectly assigned opposing directions to forces that should act in the same direction. Comparable mistakes occurred in electrostatics problems (AP Physics 2, 2017) where electric-field directions were determined without properly considering charge polarity or relative distance. Such errors often produced internally consistent calculations built upon incorrect initial vector assumptions.

\item \textbf{Qualitative and Quantitative Reasoning Errors:} LLMs occasionally demonstrated inconsistencies between qualitative explanations and quantitative analysis. Models sometimes applied formulas correctly but ignored relevant physical contributions or constraints. In AP Physics 2 (2017), involving Bernoulli's principle, several AI models failed to include the gravitational potential energy per unit volume term when estimating absolute pressure at a lower pipe section, resulting in an incomplete and incorrect application of the full Bernoulli equation. In another example (AP Physics 2, 2015), which involved light refraction through layered media (glass, liquid, and air), several models failed to correctly predict the relative positions and brightness of reflected rays. They could not provide the accurate qualitative reasoning needed to trace the rays and account for phase changes or total internal reflection.

\item \textbf{Circuit Diagram Errors:} LLMs have notable difficulty reliably identifying series and parallel relationships from schematic diagrams. They often misclassify resistor networks, miscompute equivalent resistance, and consequently mis‑rank potential differences or currents in different branches (AP Physics 2, 2022; AP Physics 1, 2016; AP Physics 2, 2015). Even when the verbal explanation gestures toward correct ideas (e.g., ``adding a parallel branch decreases equivalent resistance”), the underlying circuit topology is sometimes wrong, so the final ranking or brightness comparison fails to align with the scoring guidelines.

\item \textbf{Right-Hand Rule Errors:} The application of the right-hand rule to magnetic forces, induced currents, and magnetic fields (AP Physics 2, 2025; AP Physics 2, 2024; AP Physics 2, 2015) is a consistent source of error. AI models often ignore the sign of charge, misinterpret the direction of the magnetic field, or switch coordinate axes. These errors lead to incorrect predictions of the curvature directions of charged-particle paths or the direction of induced current in moving loops. Such mistakes highlight a broader limitation in representing three-dimensional vector relationships from static text and diagrams.
\end{enumerate}

Taken together, these error categories reveal a consistent pattern: all four LLMs perform most reliably when problems can be resolved through algebraic manipulation of well-defined equations, but performance degrades substantially when problems require accurate extraction of quantitative information from visual representations, three-dimensional spatial inference, or the coherent integration of multiple physical principles across sub-parts of a single FRQ. These findings suggest that improvements in multimodal grounding, spatial reasoning, and self-consistency checking are critical directions for enhancing the pedagogical utility of AI systems in physics education, and that human expertise remains essential for identifying and correcting the subtle reasoning failures these systems exhibit.
%
\section{Discussions}

This study provides a comprehensive, decade-spanning evaluation of four leading LLM  platforms on algebra-based AP Physics free-response examinations, offering one of the  most granular assessments to date of AI performance on authentic, multi-faceted physics reasoning tasks. While all four models achieved mean scores exceeding 82\%---a level comparable to strong human performance---the patterns underlying these scores reveal fundamental limitations that aggregate metrics alone cannot capture. This work therefore provides a longitudinal, rubric-based evaluation combining statistical performance analysis with a systematic taxonomy of physics-specific AI reasoning errors.

For AP Physics 1, the absence of statistically significant differences among models ($p = 0.141$, Kendall's $W = 0.182$) indicates that current frontier AI systems have reached broadly comparable competency in algebra-based mechanics problems. However, the low concordance coefficient suggests that this equivalence arises not from uniform mastery but from model-specific strengths that vary across exam years. Performance leadership shifted between examinations, indicating that problem framing, diagram complexity, and the balance between qualitative and quantitative reasoning interact differently with individual model architectures.

AP Physics 2 exhibited clearer differentiation. Gemini and DeepSeek achieved higher and  more consistent performance than Claude and ChatGPT, with DeepSeek demonstrating the lowest year-to-year variability ($\mathrm{CV} = 4.7\%$). The broader conceptual demands of Physics 2---including thermodynamics, optics, electromagnetism, and modern physics---appear to amplify architectural differences, particularly in tasks requiring integration of symbolic reasoning with graphical and visual information.

The qualitative error analysis provides essential explanatory context for these findings. Across all models, recurring failure modes included diagram interpretation errors, graph reading and construction errors, direction-related vector mistakes, qualitative and quantitative reasoning inconsistencies, circuit misanalysis, and right-hand rule failures.  These errors are not random---they cluster around tasks that require accurate extraction of information 
from visual representations, three-dimensional spatial inference, and the coherent application of physical principles across multiple sub-parts of a single problem. These are precisely the capabilities that distinguish genuine physics reasoning from the pattern-matching and linguistic fluency at which current LLMs excel. The finding that errors often propagated across sub-parts of an FRQ---an early diagram misreading corrupting all downstream calculations---is particularly significant for educational contexts, where partial credit depends on the integrity of the reasoning chain.

\subsection*{Recommendations for Educators and Developers}

For educators, these findings support a measured and critical integration of AI tools into physics instruction. LLMs can serve as effective supports for algebraic problem-solving, formula application, and conceptual review of well-defined topics. However, educators should caution students against over-reliance on AI for problems involving diagram interpretation, graph construction, or three-dimensional reasoning---precisely the problem types that are central to AP Physics FRQs and to authentic scientific practice. Given that errors were most likely to appear in questions requiring visual or spatial reasoning, instructors may find it pedagogically valuable to use AI-generated responses on such questions as worked examples of common misconceptions, turning model failures into teaching opportunities.

For AI developers, the error taxonomy identified in this study points to specific capability gaps that warrant targeted improvement. Enhancing multimodal grounding---the ability to accurately extract quantitative values from graphs, correctly interpret spatial relationships in diagrams, and apply the right-hand rule in three-dimensional contexts---would have the greatest impact on physics reasoning performance.  The 
frequent production of internally consistent arguments based on incorrect initial assumptions further suggests the need for improved self-consistency verification and reasoning validation mechanisms.
Chain-of-thought prompting and iterative refinement strategies, which were deliberately excluded from this study to reflect standard student usage, represent promising directions for performance improvement in future work.

Standardized assessments such as AP Physics examinations provide rigorous and reproducible benchmarks for tracking progress in AI reasoning capabilities. As LLMs continue to evolve, longitudinal evaluations of this type remain essential to distinguish genuine advances in physical reasoning from improvements tied primarily to exam familiarity or linguistic fluency. The present work establishes a baseline against which future model generations can be systematically compared, and offers a methodological template---multi-rater rubric scoring, non-parametric statistical comparison, and qualitative error taxonomy---that can be extended to other STEM assessment contexts.

In summary, current LLMs show considerable promise in physics reasoning, yet their capabilities remain incomplete. Performance is strongest in algebraic manipulation and structured problem-solving, where well-defined equations guide the solution process. Nevertheless, systematic limitations persist in visual interpretation, spatial reasoning, and multi-step inference across physical domains. Fully realizing the pedagogical potential of these systems will demand both continued advances in model architecture and thoughtful, informed integration by educators who appreciate the strengths and boundaries of AI-assisted learning.


\bibliographystyle{unsrt}
\bibliography{main}

\end{document}